\documentclass[a4paper,12pt]{article}
\usepackage[dvips]{graphics}
\usepackage{calc,pstcol,multido,pst-node,amsgen,amsmath,amsfonts,amssymb,bm}
\usepackage{rotating}
\usepackage[nolists,nomarkers,tablesfirst]{endfloat}
\usepackage{lscape,tikz}
\usetikzlibrary{positioning,shapes,arrows.meta}
\usepackage{booktabs} 

\newcommand{\ci}{\perp\!\!\!\perp}

\begin{document}
\parindent=0pt
\parskip=5pt

\begin{center}
{\LARGE \bf   Simulating from marginal structural models for hazards, cause-specific hazards and subdistribution hazards using general copulas}
\end{center}

\begin{center}
{\large \bf   Shaun R.\ Seaman$^1$}
\end{center}

$^1$ MRC Biostatistics Unit, University of Cambridge, East Forvie Building, University Forvie Site, Robinson Way, Cambridge, CB2 0SR, UK. \\
shaun.seaman@mrc-bsu.cam.ac.uk

\vspace{0.1cm}

\subsection*{Licence}

For the purpose of open access, the author has applied a Creative Commons Attribution (CC BY) licence to any Author Accepted Manuscript version arising.

\section*{Abstract}

Seaman and Keogh (Biometrical Journal 2024) proposed a method for simulating data compatible with a marginal structural model (MSM) for the hazard of a survival time outcome.
In this short report, I propose two extensions of this method.
First, Seaman and Keogh favoured the use of a Gaussian copula, because this enables the function of the confounder history through which the hazard of failure depends on confounders to be interpreted as a risk score.
Here, I describe how this interpretation can be preserved even when a non-Gaussian copula is used.
Second, I extend Seaman and Keogh's method to allow simulation of data compatible with a MSM for a cause-specific or subdistribution hazard of failure in the presence of a competing event.

\vspace{0.5cm}


\newpage

\section{Introduction}

Seaman and Keogh (2024)\cite{SeamanKeogh2024} proposed an algorithm for simulating data on a time-dependent exposure, time-dependent confounders and a failure time outcome in such a way that the data-generating mechanism is compatible with any given marginal structural model (MSM) for the hazard, e.g.\ a Cox MSM or additive hazards MSM.
This algorithm is based on the general approach of Evans and Didelez (2024)\cite{Evans2024} and overcomes limitations of previous data-simulation methods\cite{Xiao2010,Young2010,Havercroft2012,Young2014,Keogh2021}.

Seaman and Keogh's method involves two important elements.
First, the confounder history is summarised by a single function, which Seaman and Keogh call a risk score function.
Second, a copula is used to describe the association between the value of this function (called the risk score) at a given time and a latent variable that determines whether failure occurs at that time.
Seaman and Keogh stated that any copula could be used, but they promoted the use of the Gaussian copula, because it enables the risk score function to be interpreted as a function that ranks individuals with the same treatment history but different confounder histories by the magnitude of their hazard of failure.
In some circumstances, however, it may be attractive to use a non-Gaussian copula.
In this short report, I explain why, in particular, a Student-t copula might be useful, and I extend Seaman and Keogh's algorithm to enable non-Gaussian copulas to be used without losing the interpretation of the risk score function as a hazard-ranking function.

In addition, I describe how, when there is a competing event, the algorithm can be adapted to enable simulation of data compatible with a MSM for the cause-specific hazard or subdistribution hazard of failure.

\section{Set-up and notation}

\label{sect:notation}

Consider a study where individuals are observed at regular visits up to the earlier of their failure time and censoring time.
Without loss of generality, I assume that visit times are $0, 1, \ldots K$, and the administrative censoring time is $K+1$.

\begin{figure}
\begin{center}
\begin{tikzpicture}[auto, node distance=2cm, thick, node/.style={font=\sffamily\Large}]
  \node[node] (BX) {$(B, X)$};
  \node[node] (A0) [right = 2cm of BX] {$A_0$};
  \node[node] (A1) [right = 2cm of A0] {$A_1$};
  \node[node] (L0) [above left = 1.5cm and 0.5cm of A0] {$L_0$};
  \node[node] (L1) [above left = 1.5cm and 0.5cm of A1] {$L_1$};
  \node[node] (Y1) [below right = 1.5cm and 0.5cm of A0] {$Y_1$};
  \node[node] (Y2) [below right = 1.5cm and 0.5cm of A1] {$Y_2$};
  \path[every node/.style={font=\sffamily\small}]
  (BX) edge[->] node [right] {} (L0)
  (BX) edge[->] node [right] {} (A0)
  (BX) edge[->,bend right=10] node [right] {} (Y1)
  (BX) edge[->,bend left=70] node [right] {} (L1)
  (BX) edge[->,bend right=40] node [right] {} (Y2)
  (BX) edge[->,bend left=20] node [right] {} (A1)
  (A0) edge[->] node [right] {} (A1)
  (A0) edge[->] node [right] {} (L1)
  (A0) edge[->] node [right] {} (Y1)
  (A0) edge[->] node [right] {} (Y2)
  (A1) edge[->] node [right] {} (Y2)
  (L0) edge[->] node [right] {} (A0)
  (L0) edge[->] node [right] {} (A1)
  (L0) edge[->] node [right] {} (L1)
  (L0) edge[->,bend right=25] node [right] {} (Y1)
  (L0) edge[->] node [right] {} (Y2)
  (L1) edge[->] node [right] {} (A1)
  (L1) edge[->,bend left=20] node [right] {} (Y2)
  (Y1) edge[->] node [right] {} (A1)
  (Y1) edge[->] node [right] {} (Y2)
  (Y1) edge[->] node [right] {} (L1);
\end{tikzpicture}
\end{center}
\caption{Assumed causal directed acyclic graph (DAG).  For simplicity, this is shown for $K=1$.}
\label{fig:causalDAG}
\end{figure}
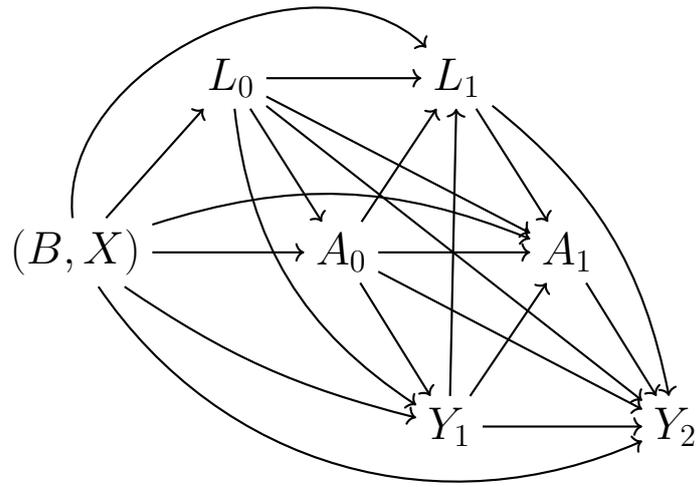

Let $A_k$ denote an individual's treatment at visit $k$ ($k=0, \ldots, K$).
This can be discrete or continuous.
Let $T$ ($T>0$) denote the individual's failure time and $Y_k=I(T \geq k)$ (for $k=1, \ldots, K+1$).
Let $L_k$ denote time-varying confounders for an individual at visit $k$ (for $k = 0,\ldots, K)$.
Let $X$ and $B$ denote two distinct vectors of baseline covariates for an individual.
Variables $X$ are baseline confounders and/or treatment effect modifiers that will be conditioned on in the MSM presented below (equation~(\ref{eq:msm})).
Variables $B$ are not conditioned on in the MSM, and can include any or all of: baseline confounders, common causes of the $L_k$ and $Y_k$ variables that are not confounders, and instrumental variables.
Either of $X$ and $B$ could be empty.
Let $\bar{A}_k = (A_0, \ldots, A_k)$ and $\bar{L}_k = (L_0, \ldots, L_k)$ denote the histories of treatment and time-dependent confounders up to visit $k$.
I assume the causal DAG shown in Figure~\ref{fig:causalDAG}.

Variables with a superscript $\bar{a}_k$ are potential variables under an intervention that sets treatments $\bar{A}_k$ at visits $0, 1, \ldots, k$ equal to $\bar{a}_k$.
For example, $Y_{k+1}^{\bar{a}_k} = 1$ if the individual would survive to visit $k+1$ if their treatments at visits $0, \ldots, k$ were set to $\bar{a}_k$, and $Y_{k+1}^{\bar{a}_k} = 0$ if the individual would fail.
We make the usual consistency assumption that $Y_k = Y_k^{\bar{A}_{k-1}}$ and $L_k = L_k^{\bar{A}_{k-1}}$.
Note that this together with the causal DAG implies sequential exchangeability, i.e.\
$
(Y_{k+1}^{\bar{A}_{k-1} \; \underline{a}_k}, \ldots, Y_{K+1}^{\bar{A}_{k-1} \; \underline{a}_k}) \ci A_k \mid X, B, \bar{L}_k, \bar{A}_{k-1}, Y_k=1
$
for all $\underline{a}_k = (a_k, \ldots, a_K)$.

\section{Seaman and Keogh's algorithm}

\label{sect:extended.algorithm}

I shall consider MSMs for discrete-time hazard of failure.
However, the methods described in this section and Section~\ref{sect:generalised.algorithm} can be extended to handle continuous time by using the approach described in Section~5 of Seaman and Keogh (2024).
Consider the following MSM for the hazard at time $k+1$ ($k=0, \ldots, K$).
\begin{equation}
P( Y_{k+1}^{\bar{a}_k} = 0 \mid X, Y_k^{\bar{a}_{k-1}} = 1 ) = g_{k+1} (\bar{a}_k, X; \beta),
\label{eq:msm}
\end{equation}
where $g_{k+1}(.)$ is a known function with parameters $\beta$.
For example, in a logistic MSM\cite{Robins2000}, $g_{k+1} (\bar{a}_k, X; \beta)$ could be chosen to be $\mbox{expit} (\beta_{k+1,0} + \beta_1^\top X + \beta_2 a_k)$.

I now describe the Basic Algorithm proposed by Seaman and Keogh for simulating data on $(X, B, L_0, A_0, Y_1, \ldots, L_K, A_K, Y_{K+1})$ such that equation~(\ref{eq:msm}) holds.

First, specify distributions $p(X)$ and $p(B \mid X)$, and $p(L_k \mid X, B, \bar{L}_{k-1}, \bar{A}_{k-1}, Y_k = 1)$ and $p(A_k \mid X, B, \bar{L}_k, \bar{A}_{k-1}, Y_k = 1)$ for each $k=0, \ldots, K$.
Also for each $k=0, \ldots, K$, specify a scalar continuous function $h_k^{\bar{a}_k} (x, b, \bar{l}_k)$ (called the risk score function) and a copula for the conditional association between risk quantile $U_{H_k^{\bar{a}_k}}$ and latent variable $U_{Y_{k+1}^{\bar{a}_k}}$ (both defined below) given $X$ and $Y_k^{\bar{a}_{k-1}} = 1$.
Seaman and Keogh assumed a Gaussian copula, but here I generalise to any pairwise copula.

A copula is a joint distribution for two random variables $U_1$ and $U_2$ whose marginal distributions are both Uniform$(0,1)$\cite{Aas2009}.
The conditional cumulative distribution function (CDF) of $U_1$ given $U_2$ implied by a copula is called the copula's h-function.
I shall denote this as $r(u_1, u_2) = P(U_1 \leq u_1 \mid U_2 = u_2)$.
For the Gaussian copula with association parameter $\rho$, the h-function is
\begin{eqnarray}
  r_{{\rm G} \rho} (u_1, u_2)
  & = &
  \Phi \left(
    \frac{ \Phi^{-1} (u_1) - \rho \Phi^{-1} (u_2) }
    { \sqrt{1 - \rho^2} }
    \right),
    \nonumber
    \\
    \label{eq:Gaussian.h}
\end{eqnarray}
where $\Phi(.)$ denotes the CDF of the standard normal distribution.
Another example of a copula is the Student-t copula with association parameter $\rho$ and $\eta$ degrees of freedom.
This has h-function
\begin{eqnarray}
  r_{{\rm S} \rho, \eta} (u_1, u_2)
  & = &
  t_{\eta+1} \left(
    \frac{ t_{\eta}^{-1} (u_1) - \rho t_{\eta}^{-1} (u_2) }
         { \sqrt{
             \frac{ \left[ \eta + \left\{ t_{\eta}^{-1} (u_2) \right\}^2 \right] (1 - \rho^2) }
                  {\eta + 1}
           }
         }
         \right),
         \nonumber \\
         \label{eq:Student.h}
\end{eqnarray}
where $t_{\eta}(.)$ denote the CDF of the Student-t distribution with $\eta$ degrees of freedom.

The steps of Seaman and Keogh's Basic Algorithm are as follows.

\vspace{.2cm}
{\it Basic Algorithm}
\begin{enumerate}
\item
  Sample $X$ from $p(X)$ and then $B$ from $p(B \mid X)$.
  Set $k=0$.
\item
  Sample $L_k$ from $p(L_k \mid X, B, \bar{L}_{k-1}, \bar{A}_{k-1}, Y_k=1)$.
\item
  Sample $A_k$ from $p(A_k \mid X, B, \bar{L}_k, \bar{A}_{k-1}, Y_k=1)$ and call the result $a_k$.
\item
  Calculate the risk score $H_k^{\bar{a}_k} = h_k^{\bar{a}_k} (X, B, \bar{L}_k)$.
\item
  Calculate the conditional quantile of this risk score given $X$ and $Y_k^{\bar{a}_{k-1}}=1$, i.e.\ $U_{H_k^{\bar{a}_k}} = F_{H_k^{\bar{a}_k}} (H_k^{\bar{a}_k} \mid X, Y_k^{\bar{a}_{k-1}}=1)$.
\item
  Set $Y_{k+1} = Y^{\bar{a}_k}_{k+1} = 0$ with probability $r \{ g_{k+1} (\bar{a}_k, X; \beta), U_{H_k^{\bar{a}_k}} \}$ and otherwise set $Y_{k+1} = Y^{\bar{a}_k}_{k+1} = 1$.
\item
  If $Y_{k+1} = 1$ and $k<K$, let $k=k+1$ and return to step 2.
\end{enumerate}
The rationale for Step 6 is as follows.
There is assumed to exist a latent variable $U_{Y_{k+1}^{\bar{a}_k}}$ that determines the value of $Y_{k+1}$.
Specifically, $Y^{\bar{a}_k}_{k+1} = 0$ if $U_{Y_{k+1}^{\bar{a}_k}} < g_{k+1} (\bar{a}_k, X; \beta)$, and $Y^{\bar{a}_k}_{k+1} = 1$ otherwise.
The conditional distribution of $U_{Y_{k+1}^{\bar{a}_k}}$ given $X$ and $Y_k^{\bar{a}_{k-1}} = 1$ is assumed to be Uniform$(0,1)$ and the conditional joint distribution of $U_{Y_{k+1}^{\bar{a}_k}}$ and $U_{H_k^{\bar{a}_k}}$ given $X$ and $Y_k^{\bar{a}_{k-1}} = 1$ is assumed to be defined by the copula.
This implies that the conditional CDF of $U_{Y_{k+1}^{\bar{a}_k}}$ given $X$, $Y_k^{\bar{a}_{k-1}} = 1$ and $U_{H_k^{\bar{a}_k}}$ evaluated at $U_{Y_{k+1}^{\bar{a}_k}} = u_1$ is given by the h-function $r(u_1, U_{H_k^{\bar{a}_k}})$ of the copula.
Because $U_{Y_{k+1}^{\bar{a}_k}}$ is Uniform$(0,1)$ given $X$ and $Y_k^{\bar{a}_{k-1}} = 1$, equation~(\ref{eq:msm}) is satisfied.

Note that Seaman and Keogh, who assumed a Gaussian copula, used two steps (their Steps 6 and 7) in place of the Step 6 given here.
For the Gaussian copula, these two steps are equivalent to Step 6 given here.

A problem with this Basic Algorithm is that Step 5 requires the CDF $F_{H_k^{\bar{a}_k}} (H_k^{\bar{a}_k} \mid X, Y_k^{\bar{a}_{k-1}}=1)$ of the risk score $H_k^{\bar{a}_k}$ given $X$ and $Y_k^{\bar{a}_{k-1}}=1$ to be known.
As this is typically unknown, Seaman and Keogh proposed using the following extended version of the Basic Algorithm, in which $F_{H_k^{\bar{a}_k}} (H_k^{\bar{a}_k} \mid X, Y_k^{\bar{a}_{k-1}}=1)$ is estimated concurrently with the simulation of the data by generating data on $m$ individuals and then discarding the data on all but the first of these $m$ individuals.
They suggested that $m=1000$ would be sufficient for most settings.

\vspace{.2cm}
{\it Extended Algorithm}
\begin{enumerate}
\item
  Sample $X_1$ from $p(X)$.
  Let $X_j = X_1$ and $I_j = j$ for $j=2, \ldots, m$.
  Set $k=0$.
\item
  For $j=1, \ldots, m$, sample $B_j$ from $p(B_j \mid X_j)$.
\item
  For $j=1, \ldots, m$, sample $L_{kj}$ from $p(L_{kj} \mid X_j, B_j, \bar{L}_{k-1,j}, \bar{A}_{k-1,1} = \bar{a}_{k-1}, Y_{kj}=1)$.
\item
  Sample $A_{k1}$ from $p(A_{k1} \mid X_1, B_1, \bar{L}_{k1}, \bar{A}_{k-1,1} = \bar{a}_{k-1}, Y_{k1}=1)$ and call the result $a_k$.
\item
  For $j=1, \ldots, m$, calculate $H_{kj} = h_k^{\bar{a}_k} (X_j, B_j, \bar{L}_{kj})$.
  \item
  For $j=1, \ldots, m$, Let $R_{kj}$ denote the rank of $H_{kj}$ among the set $\{H_{k1}, \ldots, H_{km} \}$, sample $W_{kj} \sim \mbox{Uniform} (0,1)$, and calculate $U_{H_k^{\bar{a}_k}, j} = (R_{kj} - W_{kj}) / m$.
\item
  For each of $j=1, \ldots, m$, set $Y_{k+1,j} = Y^{\bar{a}_k}_{k+1,j} = 0$ with probability $r \{ g_{k+1} (\bar{a}_k, X; \beta),$ $U_{H_k^{\bar{a}_k}, j} \}$ and otherwise set $Y_{k+1,j} = Y^{\bar{a}_k}_{k+1,j} = 1$.
\item
  If $Y_{k+1, 1} = 0$ or $k=K$, stop.
\item
  For $j=2, \ldots, m$, if $Y_{k+1,j} = 0$, randomly choose $2 \leq j^* \leq m$ such that $Y_{k+1,j^*} = 1$, and set $I_j = I_{j^*}$, $B_j = B_{j^*}$, $\bar{L}_{kj} = \bar{L}_{kj^*}$ and $Y_{k+1,j} = 1$.
\item
    Let $k=k+1$ and return to step 3.
\end{enumerate}

Again, Seaman and Keogh used three steps (their Steps 7--9) in place of the Step 7 given here; for the Gaussian copula these three steps are equivalent to Step 7 given here.
In Section E of their Supporting Materials, Seaman and Keogh also proposed a further extension of the Extended Algorithm that can improve performance of the Extended Algorithm when the failure rate is high.

\section{Gaussian versus Student-t copula}

\begin{figure}
\begin{center}
  \scalebox{0.8} {
    \includegraphics[page=1]{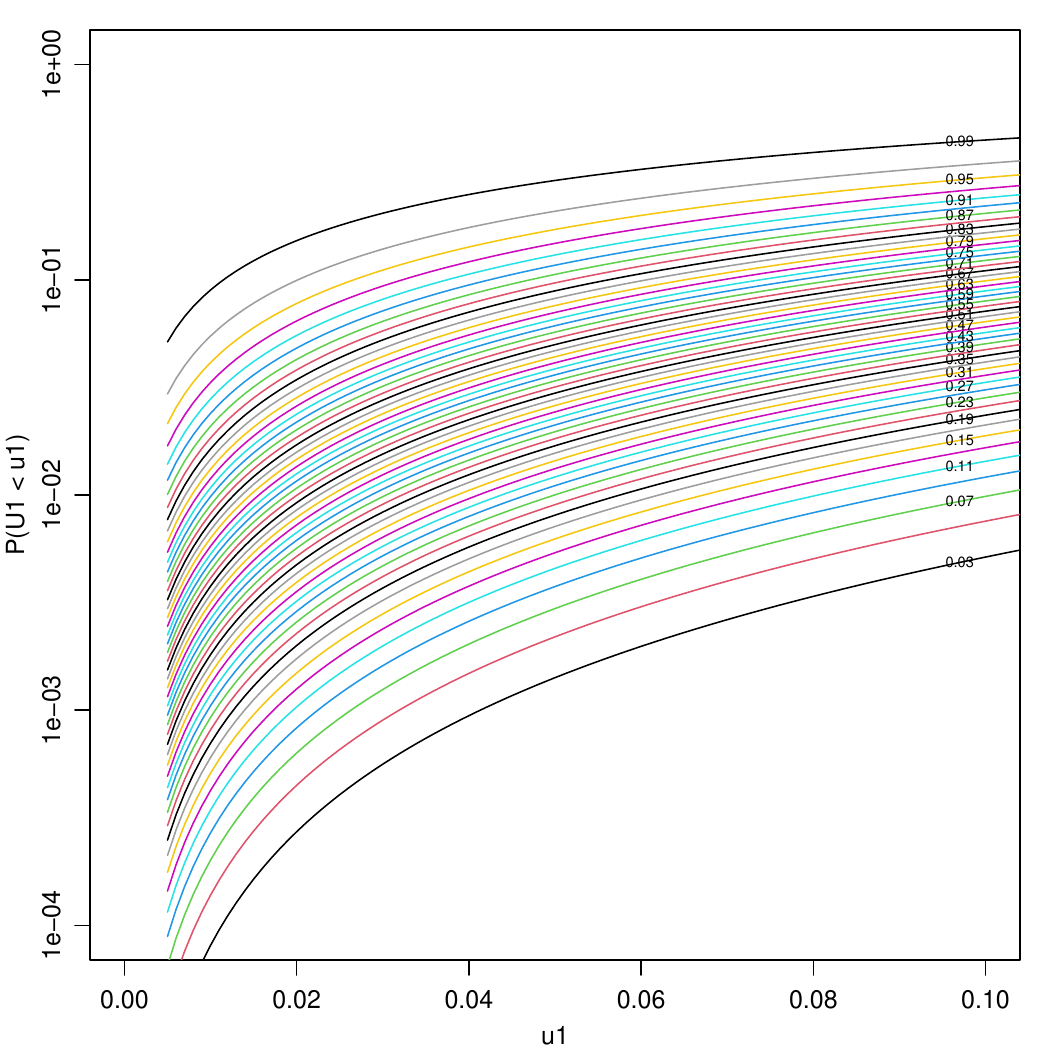}
  }
\end{center}
\caption{Probability $P(Y^{\bar{a}_k}_{k+1} = 0 \mid X, B, U_{H_k^{a_k}}, Y^{\bar{a}_{k-1}}_k=1)$ of failure (on y-axis) as a function of $g_{k+1} (\bar{a}_k, X; \beta)$ (on x-axis) and the risk quantile $U_{H_k^{a_k}}$ (the numbers on the different lines) for the Gaussian copula with $\rho = -0.5$.  Note that a log scale is used on the y-axis and that the x-axis shows only the range [0, 0.1].}
\label{fig:Gaussian.h}
\end{figure}

For any fixed value of $u_1$ and $\rho < 0$, the h-function $r_{{\rm G} \rho} (u_1, u_2)$ of the Gaussian copula is an increasing function of $u_2$.
This means that the probability that Step 6 sets $Y_{k+1}=0$ (i.e.\ failure) is an increasing function of $U_{H_k^{\bar{a}_k}}$.
This justifies the interpretation of $U_{H_k^{\bar{a}_k}}$ as the risk quantile, and hence of $H_k^{\bar{a}_k}$ as the risk score.
Figure~\ref{fig:Gaussian.h} illustrates the monotonicity of the Gaussian h-function.

\begin{figure}
\begin{center}
  \scalebox{0.8} {
    \includegraphics[page=1]{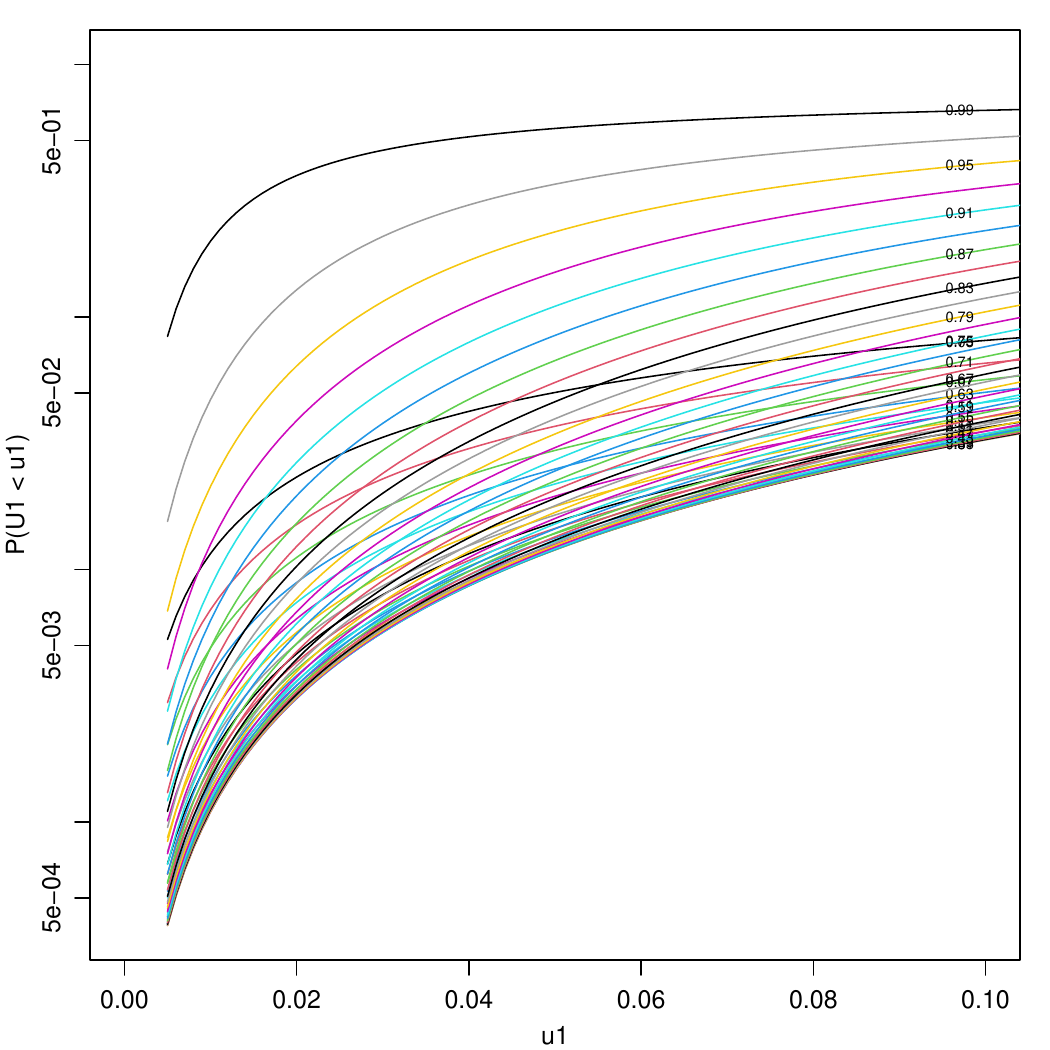}
  }
\end{center}
\caption{Probability $P(Y^{\bar{a}_k}_{k+1} = 0 \mid X, B, U_{H_k^{a_k}}, Y^{\bar{a}_{k-1}}_k=1)$ of failure (on y-axis) as a function of $g_{k+1} (\bar{a}_k, X; \beta)$ (on x-axis) and the risk quantile $U_{H_k^{a_k}}$ (the numbers on the different curves) for the Student-t copula with $\rho = -0.5$ and $\eta=2$.  Note that a log scale is used on the y-axis and that the x-axis shows only the range [0, 0.1], in order to highlight the cross-overs.}
\label{fig:Student.h}
\end{figure}

\begin{figure}
\begin{center}
  \scalebox{0.8} {
    \includegraphics[page=5]{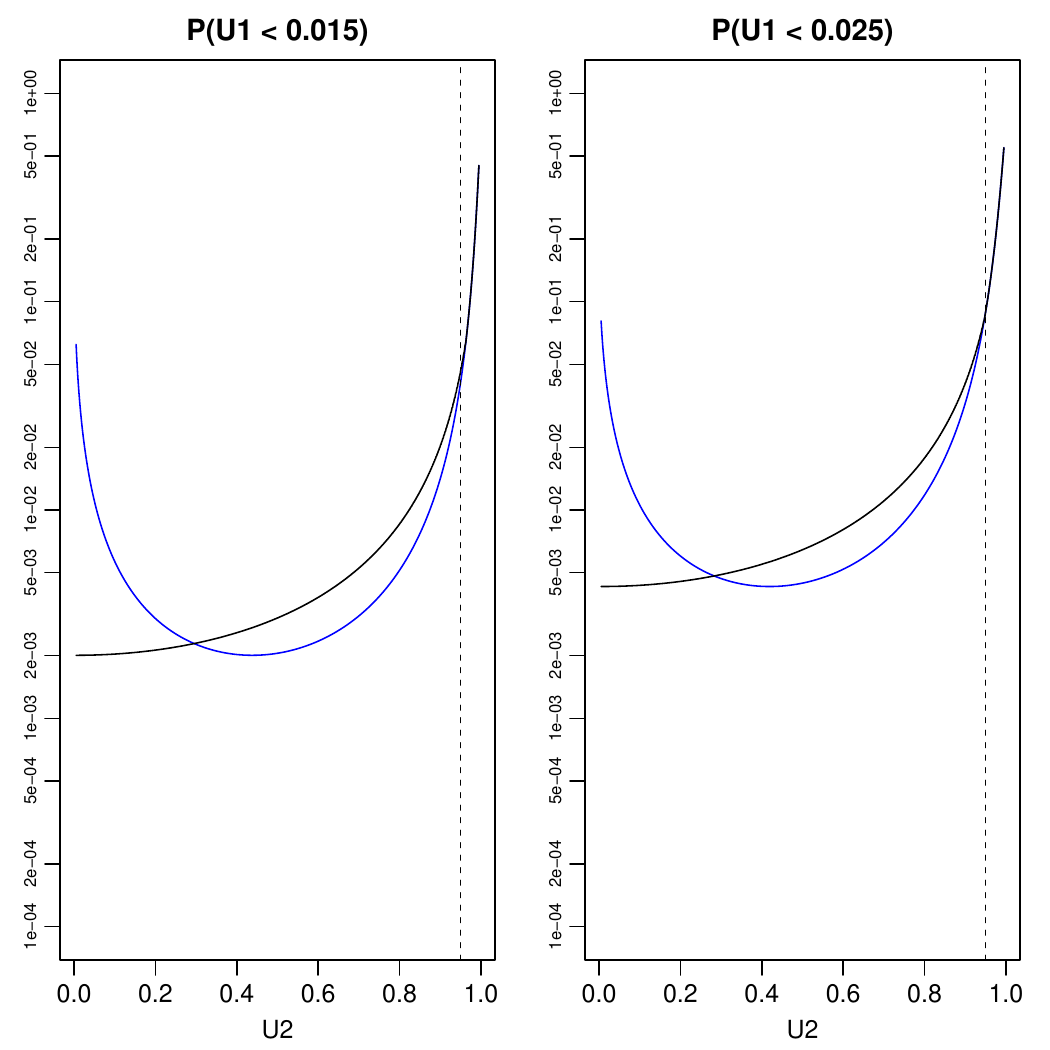}
  }
\end{center}
\caption{For Student-t copula with $\rho = -0.5$ and $\eta=2$ and for $g_{k+1} (\bar{a}_k, X; \beta)=0.015$ (left) or 0.025 (right), the blue curve is $r \{ g_{k+1} (\bar{a}_k, X; \beta), U_{H_k^{\bar{a}_k}, j} \}$ and the black curve is $q_{g_{k+1} (\bar{a}_k, X; \beta)} ( U_{H_k^{\bar{a}_k}, j} )$.  The dotted line indicates $U_{H_k^{a_k}} = 0.95$.  Note that a log scale is used on the y-axis.}
\label{fig:fixing.Student}
\end{figure}

Unlike the h-function of the Gaussian copula, the h-function $r_{{\rm S} \rho, \eta} (u_1, u_2)$ of the Student-t copula is not a monotonic function of $u_2$ for fixed $u_1$, $\rho$ and $\eta$.
H-functions of other copulas are also not necessarily monotonic functions of $u_2$.
Figure~\ref{fig:Student.h} illustrates the non-monotonicity of the Student-t h-function.
We see that many of the curves cross over other curves.
Hence, if we consider two individuals with the same value of $X$ but different values of $U_{H_k^{a_k}}$, which of these two individuals is more likely to fail at time $k+1$ may depend on the value of $g_{k+1} (\bar{a}_k, X; \beta)$.
This means that $U_{H_k^{a_k}}$ cannot easily be interpreted as a risk quantile, and hence $H_k^{\bar{a}_k}$ cannot easily be interpreted as a risk score.
Figure~\ref{fig:fixing.Student} also illustrates this feature of the Student-t copula.
The blue curves show how the hazard $P(Y^{\bar{a}_k}_{k+1} = 0 \mid X, B, U_{H_k^{a_k}}, Y^{\bar{a}_{k-1}}_k=1)$ depends on $U_{H_k^{a_k}}$ for two particular values, 0.015 and 0.025, of $g_{k+1} (\bar{a}_k, X; \beta)$ when $\rho = -0.5$ and $\eta=2$.
We see that the hazard is large not only for individuals with risk quantiles close to one but also for those with risk quantiles close to zero.
%
%

In the next section, I propose a generalisation of the Extended Algorithm that allows $U_{H_k^{a_k}}$ and $H_k^{\bar{a}_k}$ to be interpreted as the risk quantile and risk score even when a non-Gaussian copula (e.g.\ Student-t copula) is used.
First, however, I explain why one might wish to use the Student-t copula.

\begin{figure}
\begin{center}
  \scalebox{0.8} {
    \includegraphics[page=3]{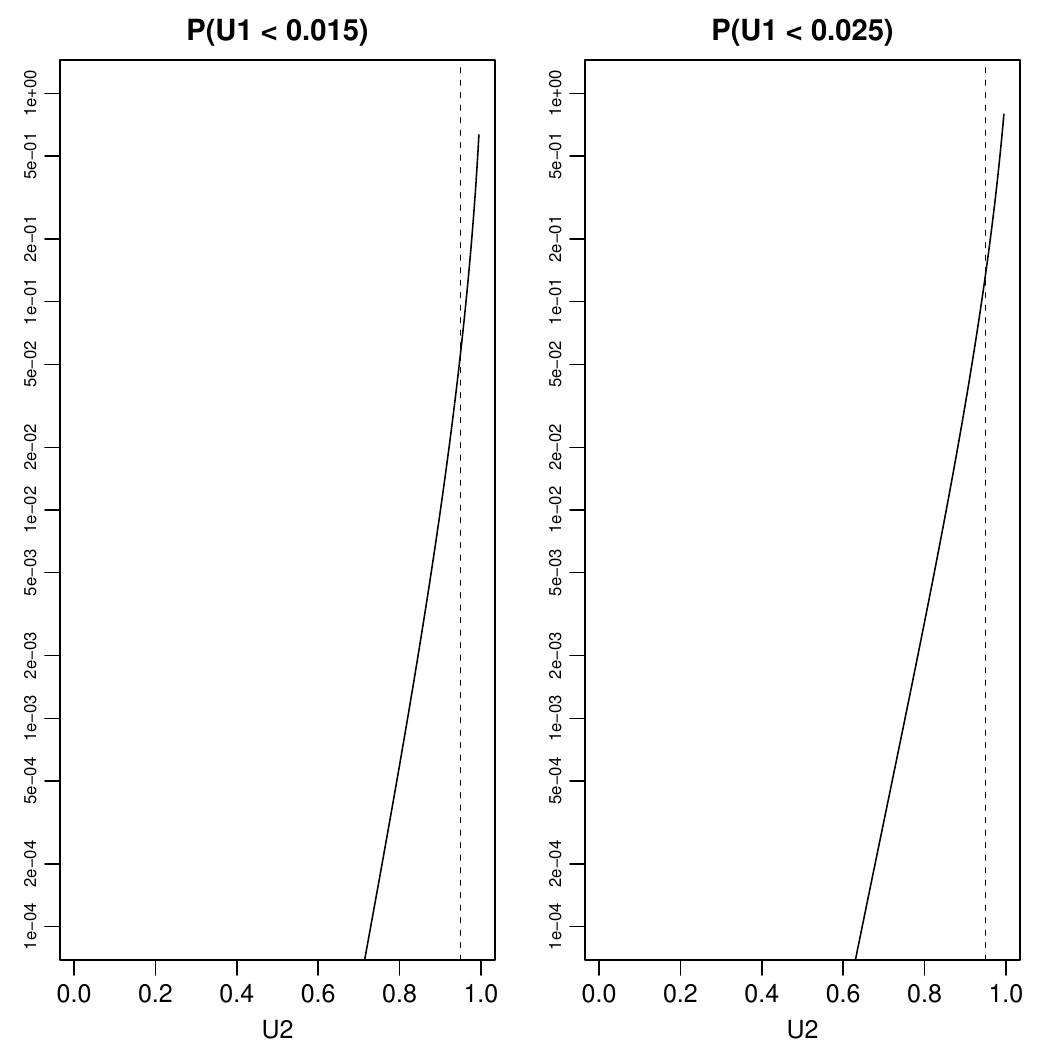}
  }
\end{center}
\caption{Probability $P(Y^{\bar{a}_k}_{k+1} = 0 \mid X, B, U_{H_k^{a_k}}, Y^{\bar{a}_{k-1}}_k=1)$ of failure (on y-axis) as a function of risk quantile $U_{H_k^{a_k}}$ (on x-axis) for two different values of marginal hazard $g_{k+1} (\bar{a}_k, X; \beta)$ (0.015 on left, 0.025 on right) when Gaussian copula with $\rho = -0.9$ is used.  The dotted line indicates $U_{H_k^{a_k}} = 0.95$.  Note that a log scale is used on the y-axis.}
\label{fig:Gaussian.h2}
\end{figure}

Suppose we wish to generate data for a scenario where the individuals with risk score in the top 5\%, i.e.\ with risk quantile $U_{H_k^{a_k}} > 0.95$, have a much higher hazard $P(Y^{\bar{a}_k}_{k+1} = 0 \mid X, B, U_{H_k^{a_k}}, Y^{\bar{a}_{k-1}}_k=1)$ than other individuals.
This can be achieved by using a Gaussian copula with a large (negative) value of $\rho$.
For example, Figure~\ref{fig:Gaussian.h2} shows how $P(Y^{\bar{a}_k}_{k+1} = 0 \mid X, B, U_{H_k^{a_k}}, Y^{\bar{a}_{k-1}}_k=1)$ depends on $U_{H_k^{a_k}}$ when the marginal hazard $g_{k+1} (\bar{a}_k, X; \beta)$ equals 0.015 (left) or 0.025 (right) and the Gaussian copula is used with $\rho = -0.9$.
Notice that this large value of $\rho$ implies that the 60\% of individuals with the lowest risk scores, i.e.\ those with $U_{H_k^{a_k}} < 0.6$, have hazards $P(Y^{\bar{a}_k}_{k+1} = 0 \mid X, B, U_{H_k^{a_k}}, Y^{\bar{a}_{k-1}}_k=1)$ that are extremely low (i.e.\ considerably less than $10^{-4}$).
If we wish these 60\% of individuals to have a less negligible hazard of failure, we need to choose $\rho$ to be closer to zero.
However, doing this will also lower the hazard of failure of the top 5\%.
By instead using the Generalised Extended Algorithm described in the next section with a Student-t copula, we can simultaneously allow the top 5\% to have very high hazards and the bottom 60\% to have non-negligible hazards.
In particular, the black curves in Figure~\ref{fig:fixing.Student} show how $P(Y^{\bar{a}_k}_{k+1} = 0 \mid X, B, U_{H_k^{a_k}}, Y^{\bar{a}_{k-1}}_k=1)$ depends on $U_{H_k^{a_k}}$ when the Generalised Extended Algorithm is used with a Student-t copula with $\rho = -0.5$ and $\eta=2$.

\section{Generalised extended algorithm for non-Gaussian copula}

\label{sect:generalised.algorithm}

The following Generalised Extended Algorithm allows the risk score function to be interpreted as a function that ranks individuals with different confounder histories according to their hazards even when a non-Gaussian copula is used.
Recall that this interpretation requires that $P(Y^{\bar{a}_k}_{k+1} = 0 \mid X, B, U_{H_k^{a_k}}, Y^{\bar{a}_{k-1}}_k=1)$ be an increasing function of the risk quantile $U_{H_k^{a_k}}$ and that this is true for the Gaussian copula (with $\rho<0$) but not necessarily for other copulas.

Before describing the Generalised Extended Algorithm, I shall explain the principle behind it.
Let $r(u_1, u_2)$ denote the h-function of any copula for marginally uniformly distributed variables $U_1$ and $U_2$.
For any fixed value of $u_1$, define the random variable $R_{u_1} = r(u_1, U_2)$.
Note that $E(R_{u_1}) = E \{ r(u_1, U_2) \} = E \{ P(U_1 \leq u_1 \mid U_2) \} = P(U_1 \leq u_1) = u_1$.
Let $q_{u_1} (v)$ denote the $v$th quantile of the marginal distribution of $R_{u_1}$,
and define random variable $R_{u_1}^* = q_{u_1} (U_2)$.
Finally, define a binary random variable $Y$ with $P(Y = 0 \mid U_2) = q_{u_1} (U_2) = R_{u_1}^*$.

For the Gaussian copula with $\rho<0$, $r (u_1, u_2)$ is an increasing function of $u_2$ (for fixed $u_1$).
Hence, since $U_2 \sim \mbox{Uniform}(0,1)$, $q_{u_1} (u_2) = r(u_1, u_2)$ and consequently $R_{u_1}^* = R_{u_1}$.
For other copulas, $r (u_1, u_2)$ may be a non-monotonic function of $u_2$, but $q_{u_1} (u_2)$ will be an increasing function of $u_2$.
Also, $E(R_{u_1}^*) = E(R_{u_1}) = u_1$.
Hence, if $Y$ is defined as a binary variable with $Y \mid U_2 \sim \mbox{Bernoulli} \left( q_{u_1} (U_2) \right)$, then $P(Y = 0 \mid U_2)$ is an increasing function of $U_2$ and $P(Y=0) = E \{ P(Y=0 \mid U_2) \} = E(R_{u_1}^*) = u_1$.
This means that if $Y=0$ represents failure, we can interpret $U_2$ as a risk quantile.

I modify the Extended Algorithm by replacing the h-function $r \{ g_{k+1} (\bar{a}_k, X; \beta),$ $U_{H_k^{\bar{a}_k}, j} \}$ by the quantile function $q_{g_{k+1} (\bar{a}_k, X; \beta)} ( U_{H_k^{\bar{a}_k}, j} )$ in Step 7.
The black curves in Figure~\ref{fig:fixing.Student} show $q_{g_{k+1} (\bar{a}_k, X; \beta)}$ $( U_{H_k^{\bar{a}_k}, j} )$ for the Student-t copula with $\rho = -0.5$ and $\eta=2$ for two values, 0.015 and 0.025, of $g_{k+1} (\bar{a}_k, X; \beta)$.
We see that using the quantile function in place of the h-function ensures that $P(Y^{\bar{a}_k}_{k+1} = 0 \mid X, B, U_{H_k^{a_k}}, Y^{\bar{a}_{k-1}}_k=1)$ is an increasing function of $U_{H_k^{a_k}}$.

In general, the quantiles $q_{g_{k+1} (\bar{a}_k, X; \beta)} ( U_{H_k^{\bar{a}_k}, j} )$ will be unknown, and so will need to be estimated.
This can be done by replacing Step 7 in the Extended Algorithm by the following step.

\vspace{0.22cm}
{\it Step 7 of Generalised Extended Algorithm}
\begin{enumerate}
  \setcounter{enumi}{6}
\item
  \begin{enumerate}
  \item
    For each $j=1, \ldots, m$, calculate $Q_{kj} = r \{ g_{k+1} (\bar{a}_k, X; \beta), U_{H_k^{\bar{a}_k}} \}$.
  \item
    Let $Q_{k(1)} < \ldots < Q_{k(m)}$ denote $Q_{k1}, \ldots, Q_{km}$ sorted into ascending order.
  \item
    Set $Y_{k+1,j} = Y^{\bar{a}_k}_{k+1,j} = 0$ with probability $Q_{k(j)}$ and otherwise set $Y_{k+1,j} = Y^{\bar{a}_k}_{k+1,j} = 1$.
  \end{enumerate}
\end{enumerate}

Steps 1--6 and 8--10 of the Generalised Extended Algorithm are identical to those of the Extended Algorithm.
Note that $Q_{k(1)}, \ldots, Q_{k(m)}$ are the estimated quantiles.
The reason for referring to this modification of the Extended Algorithm as the {\it Generalised} Extended Algorithm is that when the Gaussian copula is used, $Q_{k(j)} = Q_{kj}$ and so Steps 7(a)--7(c) are identical to Step 7 of the Extended Algorithm in that case.

\section{Empirical study}

\label{sect:simulationstudy}

To demonstrate that the Generalised Extended Algorithm does indeed generate data compatible with an assumed MSM, we repeated the empirical study described in Section F of the Supplementary Materials of Seaman and Keogh (2024).
Seaman and Keogh used the Extended Algorithm with a Gaussian copula with $\rho = -0.9$ to generate data on $n=10^6$ individuals.
In this section, we describe results obtained by using the Generalised Extended Algorithm with the Student-t copula with $\rho = -0.9$ and $\eta=2$ degrees of freedom.
As in Seaman and Keogh (see their Section 4 and Section E of Supporting Materials), the algorithm used $m=5000$ (i.e.\ 4999 matches) initially and restarted with $m=100,000$ when fewer than 10\% of the original matches remain.
However, $m=1000$ and $m=20,000$ could have been used instead, to save time.

Tables~\ref{tab:mslm.50}, \ref{tab:mslm.10} and \ref{tab:mslm.90} show the parameter estimates when the MSM was fitted with and without inverse probability of treatment weights to the simulated data on the $n=10^6$ individuals.
Table~\ref{tab:mslm.50} shows the results when the marginal probability of failure before time 10 equals 0.5, and Tables~\ref{tab:mslm.10} and~\ref{tab:mslm.90} show results when it equals 0.1 and 0.9, respectively.
We see that the parameter estimates obtained using inverse probability of treatment weighting are very close to the true values, as expected.
Also as expected, the naive estimates obtained by omitting these weights are biased, especially that for $\beta_3$, which is the causal effect of treatment.
 \begin{table}
   \begin{tabular}{crrrrr}
&& \multicolumn{2}{c}{Unweighted} & \multicolumn{2}{c}{Weighted} \\ 
Parameter     & True & Est    & SE     & Est    & SE     \\ \hline
$\beta_{00}$ &  -2.5 &  -3.026  &  0.006 &  -2.505 &  0.007  \\
$\beta_{10}$ &  -2.5 &  -3.183  &  0.006 &  -2.508 &  0.007  \\
$\beta_{20}$ &  -2.5 &  -3.213  &  0.005 &  -2.498 &  0.007  \\
$\beta_{30}$ &  -2.5 &  -3.213  &  0.005 &  -2.501 &  0.008  \\
$\beta_{40}$ &  -2.5 &  -3.205  &  0.005 &  -2.513 &  0.009  \\
$\beta_{50}$ &  -2.5 &  -3.195  &  0.006 &  -2.492 &  0.010  \\
$\beta_{60}$ &  -2.5 &  -3.178  &  0.006 &  -2.487 &  0.012  \\
$\beta_{70}$ &  -2.5 &  -3.166  &  0.006 &  -2.473 &  0.015  \\
$\beta_{80}$ &  -2.5 &  -3.163  &  0.007 &  -2.503 &  0.015  \\
$\beta_{90}$ &  -2.5 &  -3.153  &  0.007 &  -2.489 &  0.018  \\
$\beta_1$    &  0.5  &  0.436  &  0.003  &  0.505  & 0.004  \\
$\beta_2$    &  0.5  &  0.421  &  0.005  &  0.514  & 0.007  \\
$\beta_3$    & -1.0  &  0.234  &  0.005  & -1.006  & 0.007  \\
$\beta_4$    &  0.0  &  0.009  &  0.001  &  0.002  & 0.001  \\
$\beta_5$    &  0.0  &  0.010  &  0.001  &  0.005  & 0.002  \\
$\beta_6$    &  0.0  &  0.042  &  0.001  &  0.001  & 0.002
   \end{tabular}
   \caption{Estimates of parameters in MSM obtained with and without inverse probability of treatment weighting, along with estimated SEs, when the marginal probability of failure before time 10 equals 0.5.}
\label{tab:mslm.50}
\end{table}
 \begin{table}
   \begin{tabular}{crrrrr}
&& \multicolumn{2}{c}{Unweighted} & \multicolumn{2}{c}{Weighted} \\
Parameter     & True & Est    & SE     & Est    & SE     \\ \hline
$\beta_{00}$  & -4.1 &  -4.908  &  0.013 &  -4.098 &  0.018  \\
$\beta_{10}$  & -4.1 &  -5.132  &  0.013 &  -4.077 &  0.019  \\
$\beta_{20}$  & -4.1 &  -5.249  &  0.012 &  -4.101 &  0.022  \\
$\beta_{30}$  & -4.1 &  -5.320  &  0.012 &  -4.111 &  0.030  \\
$\beta_{40}$  & -4.1 &  -5.345  &  0.011 &  -4.144 &  0.031  \\
$\beta_{50}$  & -4.1 &  -5.403  &  0.012 &  -4.232 &  0.035  \\
$\beta_{60}$  & -4.1 &  -5.416  &  0.012 &  -4.101 &  0.046  \\
$\beta_{70}$  & -4.1 &  -5.453  &  0.013 &  -4.189 &  0.039  \\
$\beta_{80}$  & -4.1 &  -5.511  &  0.014 &  -4.237 &  0.046  \\
$\beta_{90}$  & -4.1 &  -5.550  &  0.016 &  -4.129 &  0.060  \\
$\beta_1$    &  0.5  &  0.424  &  0.005  &  0.499  & 0.010  \\
$\beta_2$    &  0.5  &  0.388  &  0.011  &  0.491  & 0.024  \\
$\beta_3$    & -1.0  &  0.646  &  0.012  & -1.007  & 0.019  \\
$\beta_4$    &  0.0  &  0.001  &  0.001  &  0.003  & 0.003  \\
$\beta_5$    &  0.0  &  0.006  &  0.002  &  0.004  & 0.007  \\
$\beta_6$    &  0.0  &  0.074  &  0.002  &  0.010  & 0.006
   \end{tabular}
   \caption{Estimates of parameters in MSM obtained with and without inverse probability of treatment weighting, along with estimated SEs, when the marginal probability of failure before time 10 equals 0.1.}
\label{tab:mslm.10}
\end{table}
 \begin{table}
   \begin{tabular}{crrrrr}
&& \multicolumn{2}{c}{Unweighted} & \multicolumn{2}{c}{Weighted} \\ 
Parameter     & True & Est    & SE     & Est    & SE     \\ \hline
$\beta_{00}$  & -1.2 &  -1.564  &  0.003 &  -1.205 &  0.004  \\
$\beta_{10}$  & -1.2 &  -1.654  &  0.004 &  -1.205 &  0.004  \\
$\beta_{20}$  & -1.2 &  -1.638  &  0.004 &  -1.203 &  0.004  \\
$\beta_{30}$  & -1.2 &  -1.590  &  0.004 &  -1.196 &  0.005  \\
$\beta_{40}$  & -1.2 &  -1.553  &  0.004 &  -1.209 &  0.007  \\
$\beta_{50}$  & -1.2 &  -1.513  &  0.005 &  -1.202 &  0.009  \\
$\beta_{60}$  & -1.2 &  -1.472  &  0.005 &  -1.198 &  0.011  \\
$\beta_{70}$  & -1.2 &  -1.438  &  0.006 &  -1.159 &  0.015  \\
$\beta_{80}$  & -1.2 &  -1.421  &  0.007 &  -1.166 &  0.017  \\
$\beta_{90}$  & -1.2 &  -1.406  &  0.008 &  -1.129 &  0.033  \\
$\beta_1$    &  0.5  &  0.443  &  0.002  &  0.497  & 0.003  \\
$\beta_2$    &  0.5  &  0.428  &  0.004  &  0.507  & 0.005  \\
$\beta_3$    & -1.0  & -0.029  &  0.004  & -1.000  & 0.005  \\
$\beta_4$    &  0.0  &  0.013  &  0.001  &  0.002  & 0.001  \\
$\beta_5$    &  0.0  &  0.014  &  0.001  &  0.004  & 0.002  \\
$\beta_6$    &  0.0  &  0.044  &  0.001  &  0.000  & 0.002
   \end{tabular}
   \caption{Estimates of parameters in MSM obtained with and without inverse probability of treatment weighting, along with estimated SEs, when the marginal probability of failure before time 10 equals 0.9.}
\label{tab:mslm.90}
\end{table}

The time taken to generate the data on the $n \times 3 = 3,000,000$ individuals needed for this study was three hours and 45 minutes using a single core of a Dell Latitude 5520 laptop with 4.7GHz 4-core Intel i7 processor.

\section{Competing risks}

Suppose that, in addition to the failure event of interest, there is a competing event.
I assume that failure times and competing event times are measured in discrete time.
Now, $T \in \{ 1, 2, \ldots, K+1, \infty \}$, with $T=\infty$ meaning does not fail by time $K+1$.
Similarly, let $D$ denote the time of the competing event ($D=0, 1, \ldots, K+1, \infty$), with $D=\infty$ meaning does not experience the competing event by time $K+1$.
The event time of an individual is the minimum of $T$ and $D$.
Recall that $Y_k = I(T \geq k)$, and define $Z_k = I(D \geq k)$ analogously.
Note that at least one of $T$ and $D$ must equal $\infty$, and at least one of $Y_k$ and $Z_k$ must equal 1.
Assume that the competing event causally precedes the failure event, so that $P(Y_{k+1}=0 \mid Y_k=1, Z_k=1, Z_{k+1}=0) = 0$.

\subsection{Subdistribution hazard}

Suppose we wish to generate data compatible with the following MSM for the subdistribution hazard of failure.
\begin{equation*}
P( Y_{k+1}^{\bar{a}_k} = 0 \mid X, Y_k^{\bar{a}_{k-1}} = 1 ) = g_{k+1} (\bar{a}_k, X; \beta),
\end{equation*}
where $g_{k+1}(.)$ is a known function with parameters $\beta$ (see, for example, \cite{Bekaert2010}).

This can be done using the following modified version of the Extended Algorithm of Section~\ref{sect:extended.algorithm}.
Note that Steps 1--4 are unchanged.

\vspace{0.2cm}
{\it Extended Algorithm for Subdistribution Hazard}
\begin{enumerate}
\item
  Sample $X_1$ from $p(X)$.
  Let $X_j = X_1$ and $I_j = j$ for $j=2, \ldots, m$.
  Set $k=0$.
\item
  For $j=1, \ldots, m$, sample $B_j$ from $p(B_j \mid X_j)$.
\item
  For $j=1, \ldots, m$, sample $L_{kj}$ from $p(L_{kj} \mid X_j, B_j, \bar{L}_{k-1,j}, \bar{A}_{k-1,1} = \bar{a}_{k-1}, Y_{kj}=1)$.
\item
  Sample $A_{k1}$ from $p(A_{k1} \mid X_1, B_1, \bar{L}_{k1}, \bar{A}_{k-1,1} = \bar{a}_{k-1}, Y_{k1}=1)$ and call the result $a_k$.
\item
  For each $j=1, \ldots, m$ such that $Z_{kj}=1$, sample $Z_{k+1,j} \sim \mbox{Bernoulli} \left( s_{k+1} (X_j, B_j, \bar{L}_{kj}, \bar{a}_k) \right)$ for some choice of function $s_{k+1} (.)$.
  If $Z_{k+1,1} = 0$, then stop.
\item
  For $j=1, \ldots, m$ such that $Z_{k+1,j}=1$, calculate $H_{kj} = h_k^{\bar{a}_k} (X_j, B_j, \bar{L}_{kj})$.
  \item
  For $j=1, \ldots, m$ such that $Z_{k+1,j}=1$, let $R_{kj}$ denote the rank of $H_{kj}$ among the set $\{H_{kj}: j=1, \ldots, m \mbox{ such that } Z_{k+1,j}=1 \}$, sample $W_{kj} \sim \mbox{Uniform} (0,1)$, and calculate $U_{H_k^{\bar{a}_k}, j} = (R_{kj} - W_{kj}) / \sum_{l=1}^m Z_{k+1,l}$.
\item
  For each of $j=1, \ldots, m$ such that $Z_{k+1,j}=1$, set $Y_{k+1,j} = Y^{\bar{a}_k}_{k+1,j} = 0$ with probability $r \{ g_{k+1} (\bar{a}_k, X; \beta),$ $U_{H_k^{\bar{a}_k}, j} \} / \left( 1 - \left. \sum_{l=1}^m Z_{k+1,l} \; \right/ m \right)$ and otherwise set $Y_{k+1,j} = Y^{\bar{a}_k}_{k+1,j} = 1$.
\item
  If $Y_{k+1, 1} = 0$ or $k=K$, stop.
\item
  For $j=2, \ldots, m$, if $Y_{k+1,j} = 0$, randomly choose $2 \leq j^* \leq m$ such that $Y_{k+1,j^*} = 1$, and set $I_j = I_{j^*}$, $B_j = B_{j^*}$, $\bar{L}_{kj} = \bar{L}_{kj^*}$, $\bar{Z}_{k+1,j} = \bar{Z}_{k+1,j^*}$ and $Y_{k+1,j} = 1$.
\item
    Let $k=k+1$ and return to Step 3.
\end{enumerate}
Note the division of $r \{ g_{k+1} (\bar{a}_k, X; \beta),$ $U_{H_k^{\bar{a}_k}, j} \}$ by $( 1 - \left. \sum_{l=1}^m Z_{k+1,l} \; \right/m )$ in Step 8.
This is done because individuals who have experienced the competing event by time $k+1$ cannot fail at time $k+1$ and the risk set on which the subdistribution hazard at time $k+1$ is based includes these individuals.

If a non-Gaussian copula is used, the approach described in Section~\ref{sect:generalised.algorithm} can be used to ensure that the hazard of failure is an increasing function of the risk quantile $U_{H_k^{\bar{a}_k}}$.

\subsection{Cause-specific hazard}

\label{sect:cause-specific}

Suppose instead that we wish to generate data compatible with the following MSM for the cause-specific hazard of failure.
\begin{equation*}
P( Y_{k+1}^{\bar{a}_k} = 0 \mid X, Y_k^{\bar{a}_{k-1}} = 1, Z_{k+1}^{\bar{a}_k} = 1 ) = g_{k+1} (\bar{a}_k, X; \beta),
\end{equation*}
where $g_{k+1}(.)$ is a known function with parameters $\beta$ (see, for example, \cite{Moodie2014}).
I shall assume that we also have a MSM for the cause-specific hazard, $P( Z_{k+1}^{\bar{a}_k} = 0 \mid X, Z_k^{\bar{a}_{k-1}} = 1, Y_k^{\bar{a}_{k-1}} = 1 )$, of the competing event and a risk score function and copula for this competing event.
However, this is not necessary.

Data can be simulated using the Extended Algorithm of Section~\ref{sect:extended.algorithm} but with the following modifications.
After Steps 1--4, the risk quantiles for the competing event are calculated using the risk score function for the competing event, and the values of $Z_{k+1,1}, \ldots, Z_{k+1,m}$ are generated.
This procedure is like Steps 5--7, but for $Z_{k+1,j}$ rather than $Y_{k+1,j}$.
Alternatively, $Z_{k+1,j}$ can be generated conditional on $(X_j, B_j, \bar{L}_{kj}, \bar{A}_{kj})$ and $Z_{kj}^{\bar{a}_{k-1}} = Y_{kj}^{\bar{a}_{k-1}} = 1$ in any way that is desired.
Stop if $Z_{k+1,1} = 0$.
Otherwise, the original Steps 5--7 are then executed, but with two changes.
First, only those of the $m$ individuals with $Z_{k+1, j}=0$ are used to calculate the ranks of $H_{kj}$ and the risk quantiles $U_{H_k^{\bar{a}_k}, j}$ in Steps 5 and 6.
Second, all those individuals with $Z_{k+1,j} = 0$ have $Y_{k+1,j}$ set to 1 in Step 7.
In Step 9, if $Y_{k+1,j} = 0$ or $Z_{k+1,j} = 0$, randomly choose $2 \leq j^* \leq m$ such that $Y_{k+1,j} = Z_{k+1,j} = 1$.

Again, if a non-Gaussian copula is used, the approach described in Section~\ref{sect:generalised.algorithm} can be used.

\section{Discussion}

\label{sect:discussion}

In order to use the Generalised Extended Algorithm, an explicit expression for the h-function of the copula needs to be known.
This expression is available for many copulas, including the Gaussian, Student-t, Clayton, Gumbel, Frank and Joe copulas\cite{Aas2009,Schepsmeier2014}.

Note that in some circumstances we may be happy with the non-monotonicity of the Student-t or other copula: specifically, if we believe that both high and low values of the risk score are associated with failure (e.g.\ very overweight and very underweight individuals may be at more risk).

Recently, Lin et al.\ (2025)\cite{Lin2025} proposed an alternative to Seaman and Keogh's (2024) algorithm.
Lin et al.\ describe the association between the time-dependent confounders and failure by using multiple pair copulas.
Their approach has two major advantages over the approach of Seaman and Keogh.
First, it is considerably more computationally efficient, because it avoids the need to simulate data on $m=1000$ individuals and then discard the data on all but one of these individuals, which Seaman and Keogh's algorithm does in order to estimate the CDF of the risk score.
Second, it avoids the need for Monte Carlo estimation of a CDF, being instead an exact algorithm.
However, Lin et al.'s approach has some disadvantages.
First, the way that the association between failure and the time-dependent confounders is described through a sequence of conditional associations given `earlier' confounders in some arbitrary ordering of the confounders at each time point arguably makes it difficult to interpret exactly how the hazard of failure depends on those confounders.
Furthermore, this difficulty may be compounded when non-Gaussian copulas are used, due to the resulting non-monotonic relation between confounder values and the hazard noted here in Section~\ref{sect:generalised.algorithm}.
Seaman and Keogh's risk function may be easier to understand, especially as the concept of a risk score is familiar in Medical Statistics, being commonly used for risk prediction; and I have shown in Section~\ref{sect:generalised.algorithm} of this current paper how the issue of non-monotonicity can be overcome.
Second, as Lin et al.\ note, their approach makes it difficult to generate data such that the hazard of failure at any time point depends only on the most recent values of the confounders (although they do briefly suggest that this should be possible if all the pair copulas are Gaussian).
This sort of Markov assumption is easy to ensure with Seaman and Keogh's approach.
Third, in this current paper, I have described simple adaptations to Seaman and Keogh's algorithm that make it simulate data compatible with a MSM for a cause-specific or subdistribution hazard.
Lin et al.\ comment that extending their approach to do this is `non-trivial'.

\subsubsection*{Licence}

For the purpose of open access, the author has applied a Creative Commons Attribution (CC BY) licence to any Author Accepted Manuscript version arising.

\subsubsection*{ORCID}

Shaun R. Seaman https://orcid.org/0000-0003-3726-5937 \\

\end{document}